# Vacuum Attraction, Friction and Heating of Nanoparticles moving nearby a heated Surface

G.V.Dedkov, A.A.Kyasov

Nanoscale Physics Group, Kabardino –Balkarian State University, Nalchik, 360004,

Russian Federation; e –mail: gv_dedkov@mail.ru

We review the fluctuation electromagnetic theory of attraction, friction and heating of neutral non –magnetic nanoparticles moving with constant velocity in close vicinity to the solid surface. The theory is based on exact solution of the relativistic problem of fluctuation electromagnetic interaction in configuration sphere –plane in dipole approximation.

## 1. Introduction

Vacuum attraction, friction and heating of neutral non –magnetic particles moving nearby the solid surface are the most accessible effects of electromagnetic fluctuations. Often, the corresponding problems are considered separately each other despite their common physical origin. Thus, vacuum attraction force between the bodies caused by fluctuation electromagnetic interaction (FEI) is known as van –der –Waals (Casimir) force [1,2], vacuum friction force is a dissipative component of this force [3,4], and heating (cooling) of a body in thermal field of another body is a manifestation of radiation heat transfer mediated by evanescent electromagnetic waves or by propagating waves [5,6]. In the latter case massive black bodies emit thermal electromagnetic radiation according to the Stefan –Boltzman law. For particles with dimensions $D < \lambda_T$ ($\lambda_T$ is the wave length of the thermal radiation), or being situated nearby a heated surface at a distance $d < \lambda_T$, the radiation heat transfer via the evanescent waves plays a dominating role.

To date, the field of FEI has a lot of puzzles, especially in the friction problem (see, for instance, discussions in [7-9]). And even in the case of conservative van –der –Waals and Casimir forces, some new issues have attracted enormous efforts (see, for example, [10,11] and references there in). Quite recently, some aspects of the aforementioned problems have been reviewed in Refs. [9,12], but the authors did not give necessary attention to several important results obtained by us in [7,8,13-16]. Thus, the criticism raised in [9,17] (see also references there in), referring to our earlier publications [18,19] is now of only historical value. Therefore, the basic aim of this paper is to picturing our present day understanding of the subject.

We treat all the problems of attraction, friction and heating of a small particle using a closed relativistic theory of FEI developed in [13,14]. This allows to reproduce not only all existing results being addressed to the involved nonrelativistic statement of the problem [4,7-9, 12, 20-23], but also, to get a lot of new results, greatly improving general understanding of the subject. Our approach is based on relativistic and fluctuating electrodynamics, using local approximation to the dielectric and magnetic permeability functions of contacting materials. A moving particle has both electric and magnetic fluctuating dipole moments in its rest frame. A fluctuating magnetic moment appreciably influences FEI and appears even on a resting nonmagnetic object due to stochastic Foucault currents being induced by random external magnetic fields penetrating inside a volume of an object having non –zero magnetic polarizability [24]. An advantage of the relativistic framework is due to an automatic incorporation of retardation, magnetic and thermal effects which manifest strong and nontrivial interplay.

Due to the lack of space, we leave aside another approach to the problems of attraction, friction and heating, which is related with the geometry of two parallel plates in rest or in relative motion. The necessary information can be found in Refs. [9,10,12, 30].

The paper is organized as follows :

In sect.2 we summarize the problem statement, main definitions and assumptions relevant to further theoretical consideration.

In sect.3 we obtain an equivalent form for the Lorentz electromagnetic force and heating rate of a particle embedded into the fluctuating electromagnetic field.

In sect.4 we derive relativistic formulas for the conservative and dissipative forces on a moving particle and its heating rate.

In sect.5 we get the expression for an attractive nonrelativistic force (Casimir force)

In sect.6 we obtain the expression for the retarded vacuum friction force up to the terms linear in velocity .

In sect.7 we obtain the expression for the heat flux between a resting particle and the surface.

In sect. 8 we consider application of the results to the problems of damping motion of an AFM tip, friction and heating of nano- and micro –sized particles in the nonretarded and retarded regimes of interaction with metallic surfaces.

Finally, in sect.9 we present concluding remarks.

## 2. Problem statement

Let us consider a small–sized neutral spherical particle of radius $R$ that moves in vacuum parallel to the boundary of a semi–infinite medium at a distance $z_0$ from the boundary (Fig.1). The particle velocity $V$ is directed along the $x$–axis and is arbitrary. The half–space $z<0$ is filled by homogeneous and isotropic medium characterized by complex dielectric permittivity $\varepsilon$ and permeability $\mu$ depending on the frequency $\omega$. It is assumed that the particle has the temperature $T_1$ and the medium and surrounding electromagnetic vacuum background - $T_2$. The global system is out of thermal equilibrium, but in a stationary regime. Also, it is assumed that the particle is non–magnetic in its rest frame and has the electric and magnetic polarizabilities $\alpha_e(\omega)$ and $\alpha_m(\omega)$. The laboratory system $K$ is related to the resting surface and the coordinate system $K'$ is related to the moving particle. In dipole approximation $z_0 \gg R$, the particle can be considered as a point fluctuating dipole with random electric and magnetic moments $\mathbf{d}(t), \mathbf{m}(t)$. Our aim is to determine the force of FEI applied to a moving particle and its heating rate in the process of radiation heat transfer.

## 3. Fluctuation electromagnetic force and heating rate: general relations

First of all, we obtain several relationships corresponding to the fluctuation electromagnetic force and rate of radiation heating of a particle embedded into electromagnetic field. In the laboratory $K$–system (Fig.1), vectors of electric and magnetic polarization of a moving particle are

$$\mathbf{P}(\mathbf{r},t) = \mathbf{d}(t)\delta(\mathbf{r}-\mathbf{V}t) \quad (1)$$

$$\mathbf{M}(\mathbf{r},t) = \mathbf{m}(t)\delta(\mathbf{r}-\mathbf{V}t) \quad (2)$$

Using (1), (2), the Maxwell equations $\text{rot}\mathbf{E} = -\frac{1}{c}\frac{\partial \mathbf{B}}{\partial t}, \text{div}\mathbf{B} = 0$ and general relations for the charge and current densities, $\rho = -\text{div}\mathbf{P}, \mathbf{j} = \frac{\partial \mathbf{P}}{\partial t} + c \cdot \text{rot}\mathbf{M}$, the expression for an averaged Lorentz force can be represented in the form [16]

$$\mathbf{F} = \int \langle \rho \mathbf{E} \rangle d^3r + \frac{1}{c}\int \langle \mathbf{j} \times \mathbf{B} \rangle d^3r = \\ = \langle \nabla(\mathbf{dE}+\mathbf{mB}) \rangle + \frac{1}{c}\left\langle \frac{\partial}{\partial t}(\mathbf{d}\times\mathbf{B}) \right\rangle + \frac{1}{c}\langle (\mathbf{V}\nabla)(\mathbf{d}\times\mathbf{B}) \rangle = \langle \nabla(\mathbf{dE}+\mathbf{mB}) \rangle + \frac{1}{c}\left\langle \frac{d}{dt}(\mathbf{d}\times\mathbf{B}) \right\rangle \quad (3)$$

where the angular brackets $\langle...\rangle$ imply total quantum and statistical averaging and the integrals are taken over the particle volume. If the electromagnetic field is non fluctuating, all the terms in (3) must be taken without averaging. In the case of stationary fluctuations of electromagnetic field, the last term of (3) equals zero and we get

$$\mathbf{F} = \langle \nabla(\mathbf{dE} + \mathbf{mB}) \rangle \tag{4}$$

The heating rate of a particle in the $K'$ system is, obviously, given by the dissipation integral

$$\frac{dQ'}{dt'} = \int \langle \mathbf{j'} \cdot \mathbf{E'} \rangle d^3 r' \tag{5}$$

Furthermore, making use of the relativistic transformations for the current density, electric field and volume in the integrand (5), we obtain [14]

$$\int \langle \mathbf{j'} \cdot \mathbf{E'} \rangle d^3 r' = \gamma^2 \left( \int \langle \mathbf{j} \cdot \mathbf{E} \rangle d^3 r - \mathbf{F} \cdot \mathbf{V} \right) \tag{6}$$

where $\gamma = (1-\beta^2)^{-1/2}$ is the Lorentz –factor, $\beta = V/c$ and $\mathbf{F}$ is given by (4). According to the Planck's formulation of relativistic thermodynamics, $dQ'/dt' = \gamma^2 dQ/dt$ [26], and Eq.(6) yields

$$\frac{dQ}{dt} = \int \langle \mathbf{j} \cdot \mathbf{E} \rangle d^3 r - \mathbf{F} \cdot \mathbf{V} \tag{7}$$

Formula (7) has clear physical meaning: the energy dissipated is transformed into heat and work of the fluctuation electromagnetic force performed over the particle. The dissipation integral in (7) can be rewritten by analogy with the Lorentz force (the points above vectors $\mathbf{d}$ and $\mathbf{m}$ denote the time differentiation):

$$\int \mathbf{j} \cdot \mathbf{E} d^3 r = \int \frac{\partial \mathbf{P}}{\partial t} \cdot \mathbf{E} d^3 r + c \int \text{rot} \mathbf{M} \cdot \mathbf{E} d^3 r =$$

$$= (\dot{\mathbf{d}} \cdot \mathbf{E} + \dot{\mathbf{m}} \cdot \mathbf{B}) + \mathbf{V} \cdot \nabla(\mathbf{d} \cdot \mathbf{E} + \mathbf{m} \cdot \mathbf{B}) - \frac{d}{dt}(\mathbf{m} \cdot \mathbf{B}) = \tag{8}$$

$$= (\dot{\mathbf{d}} \cdot \mathbf{E} + \dot{\mathbf{m}} \cdot \mathbf{B}) + \mathbf{V} \cdot \mathbf{F} - \frac{d}{dt}(\mathbf{m} \cdot \mathbf{B}) - \frac{\mathbf{V}}{c} \frac{d}{dt} \mathbf{d} \times \mathbf{B}$$

Again, assuming the case of stationary fluctuations, the last two terms in (8) are equal to zero after averaging, and from (7) and (8) we finally get

$$dQ/dt = \langle \dot{\mathbf{d}} \cdot \mathbf{E} + \dot{\mathbf{m}} \cdot \mathbf{B} \rangle \tag{9}$$

Moreover, it is not difficult to show that in the $K'$ system the form of Eq.(9) does not change: $dQ'/dt' = \langle \dot{\mathbf{d}}' \cdot \mathbf{E}' + \dot{\mathbf{m}}' \cdot \mathbf{B}' \rangle$. Eqs. (4) and (9) are the primary ones in further calculations.

## 4. General relativistic expressions for conservative and dissipative components of the interaction force and heating rate

Following our method [7,8,13,14], the quantities in the right hand sides of (4) and (9) have to be conveniently written as the products of spontaneous and induced terms, i.e.

$$\mathbf{F} = \left\langle \nabla \left( \mathbf{d}^{sp} \cdot \mathbf{E}^{ind} + \mathbf{d}^{ind} \cdot \mathbf{E}^{sp} + \mathbf{m}^{sp} \cdot \mathbf{B}^{ind} + \mathbf{m}^{ind} \cdot \mathbf{B}^{sp} \right) \right\rangle \tag{10}$$

$$dQ/dt = \left\langle \dot{\mathbf{d}}^{sp} \cdot \mathbf{E}^{ind} + \dot{\mathbf{d}}^{ind} \cdot \mathbf{E}^{sp} + \dot{\mathbf{m}}^{sp} \cdot \mathbf{B}^{ind} + \dot{\mathbf{m}}^{ind} \cdot \mathbf{B}^{sp} \right\rangle \tag{11}$$

The projections of the force $F_z$ and $F_x$ onto directions $z, x$ of the system $K$ correspond to the attractive conservative (Casimir) and dissipative (frictional) force. Important technical details of these calculations can be found in [13,14], thus we only briefly touch upon the main ideas. So, all the quantities in the right hand side of (10), (11) have to be Fourier transformed over the time and space variables $t, x, y$. The induced field components $\mathbf{E}^{ind}, \mathbf{B}^{ind}$ must satisfy the necessary boundary conditions at $z = 0$ and Maxwell equations, which contain the extraneous fluctuating currents induced by the fluctuating moments $\mathbf{d}^{sp}, \mathbf{m}^{sp}$. The involved correlators of dipole moments are then calculated using the fluctuation–dissipation theorem [2]. The induced dipole moments $\mathbf{d}^{ind}, \mathbf{m}^{ind}$ have to be expressed through the fluctuating fields $\mathbf{E}^{sp}, \mathbf{B}^{sp}$, being composed of the sum of the contributions from the surface and vacuum background. The induced dipole moments are related with electric and magnetic polarizabilities of the paricle and the fields $\mathbf{E}^{sp}, \mathbf{B}^{sp}$ via linear integral relations. Arising correlators are calculated with the help of the components of the retarded photon Green function [2]. Finally, after some tedious but straightforward algebra, the needed components of the interaction force and rate of the particle heating are given by

$$F_z = -\frac{\hbar \gamma}{\pi^2} \iiint_{k>\omega/c} d\omega dk_x dk_y \exp(-2q_0 z) \cdot$$

$$\cdot \begin{Bmatrix} \coth\frac{\hbar \gamma \omega^+}{2k_B T_1}\left[\operatorname{Re} R_e^+(\omega,\mathbf{k}) \cdot \alpha_e''(\gamma\omega^+) + \operatorname{Re} R_m^+(\omega,\mathbf{k})\alpha_m''(\gamma\omega^+)\right] + \\ + \coth\frac{\hbar \gamma \omega^-}{2k_B T_1}\left[\operatorname{Re} R_e^-(\omega,\mathbf{k}) \cdot \alpha_e''(\gamma\omega^-) + \operatorname{Re} R_m^-(\omega,\mathbf{k})\alpha_m''(\gamma\omega^-)\right] + \\ + \coth\frac{\hbar \omega}{2k_B T_2}\left[\operatorname{Im} R_e^+(\omega,\mathbf{k}) \cdot \alpha_e'(\gamma\omega^+) + \operatorname{Im} R_m^+(\omega,\mathbf{k})\alpha_m'(\gamma\omega^+)\right] + \\ + \coth\frac{\hbar \omega}{2k_B T_2}\left[\operatorname{Im} R_e^-(\omega,\mathbf{k}) \cdot \alpha_e'(\gamma\omega^-) + \operatorname{Im} R_m^-(\omega,\mathbf{k})\alpha_m'(\gamma\omega^-)\right] \end{Bmatrix}$$

$$-\frac{\hbar \gamma}{\pi^2} \iiint_{k<\omega/c} d\omega dk_x dk_y \cos(2\tilde{q}_0 z) \cdot \{R_e^\pm, R_m^\pm \to \tilde{R}_e^\pm, \tilde{R}_m^\pm\} -$$

$$-\frac{\hbar \gamma}{\pi^2} \iiint_{k<\omega/c} d\omega dk_x dk_y \sin(-2\tilde{q}_0 z) \cdot$$

$$\cdot \begin{Bmatrix} \coth\frac{\hbar \gamma \omega^+}{2k_B T_1}\left[\operatorname{Im} \tilde{R}_e^+(\omega,\mathbf{k}) \cdot \alpha_e''(\gamma\omega^+) + \operatorname{Im} \tilde{R}_m^+(\omega,\mathbf{k})\alpha_m''(\gamma\omega^+)\right] + \\ + \coth\frac{\hbar \gamma \omega^-}{2k_B T_1}\left[\operatorname{Im} \tilde{R}_e^-(\omega,\mathbf{k}) \cdot \alpha_e''(\gamma\omega^-) + \operatorname{Im} \tilde{R}_m^-(\omega,\mathbf{k})\alpha_m''(\gamma\omega^-)\right] + \\ - \coth\frac{\hbar \omega}{2k_B T_2}\left[\operatorname{Re} \tilde{R}_e^+(\omega,\mathbf{k}) \cdot \alpha_e'(\gamma\omega^+) + \operatorname{Re} \tilde{R}_m^+(\omega,\mathbf{k})\alpha_m'(\gamma\omega^+)\right] + \\ - \coth\frac{\hbar \omega}{2k_B T_2}\left[\operatorname{Re} \tilde{R}_e^-(\omega,\mathbf{k}) \cdot \alpha_e'(\gamma\omega^-) + \operatorname{Re} \tilde{R}_m^-(\omega,\mathbf{k})\alpha_m'(\gamma\omega^-)\right] \end{Bmatrix} \quad (12)$$

$$F_x = -\frac{\hbar \gamma}{\pi c^4} \int_0^\infty d\omega \omega^4 \int_{-1}^1 dx\, x(1+\beta x)^2 [\alpha_e''(\omega_1) + \alpha_m''(\omega_1)] W(\omega/T_2, \omega_1/T_1) -$$

$$-\frac{\hbar \gamma}{\pi^2} \iiint_{k>\omega/c} d\omega d^2k\, k_x q_0^{-1} \exp(-2q_0 z) \cdot$$

$$\cdot \begin{Bmatrix} W(\omega/T_2, \omega^+\gamma/T_1)[\alpha_e''(\omega^+\gamma)\operatorname{Im} R_e^+(\omega,\mathbf{k}) + \alpha_m''(\omega^+\gamma)\operatorname{Im} R_m^+(\omega,\mathbf{k})] - \\ - W(\omega/T_2, \omega^-\gamma/T_1)[\alpha_e''(\omega^-\gamma)\operatorname{Im} R_e^-(\omega,\mathbf{k}) + \alpha_m''(\omega^-\gamma)\operatorname{Im} R_m^-(\omega,\mathbf{k})] \end{Bmatrix} -$$

$$-\frac{\hbar \gamma}{\pi^2} \iiint_{k<\omega/c} d\omega d^2k\, k_x \tilde{q}_0^{-1} (-\sin(2\tilde{q}_0 z))\{R_e^\pm, R_m^\pm \to \tilde{R}_e^\pm, \tilde{R}_m^\pm\} -$$

$$-\frac{\hbar \gamma}{\pi^2} \iiint_{k<\omega/c} d\omega d^2k\, k_x \tilde{q}_0^{-1} \cos(2\tilde{q}_0 z) \cdot$$

$$\cdot \begin{Bmatrix} W(\omega/T_2, \omega^+\gamma/T_1)[\alpha_e''(\omega^+\gamma)\operatorname{Re} \tilde{R}_e^+(\omega,\mathbf{k}) + \alpha_m''(\omega^+\gamma)\operatorname{Re} \tilde{R}_m^+(\omega,\mathbf{k}] - \\ - W(\omega/T_2, \omega^-\gamma/T_1)[\alpha_e''(\omega^-\gamma)\operatorname{Re} \tilde{R}_e^-(\omega,\mathbf{k}) + \alpha_m''(\omega^-\gamma)\operatorname{Re} \tilde{R}_m^+(\omega,\mathbf{k}] \end{Bmatrix} \quad (13)$$

$$\dot{Q} = \frac{\hbar \gamma}{\pi c^3} \int_0^\infty d\omega \omega^4 \int_{-1}^1 dx (1+\beta x)^3 [\alpha_e''(\omega_1) + \alpha_m''(\omega_1)] W(\omega/T_2, \omega_1/T_1) -$$

$$+ \frac{\hbar \gamma}{\pi^2} \iiint_{k>\omega/c} d\omega d^2 k \, q_0^{-1} \exp(-2 q_0 z) \cdot$$

$$\cdot \begin{cases} \omega^+ \cdot W(\omega/T_2, \omega^+ \gamma/T_1) [\alpha_e''(\omega^+ \gamma) \operatorname{Im} R_e^+(\omega,\mathbf{k}) + \alpha_m''(\omega^+ \gamma) \operatorname{Im} R_m^+(\omega,\mathbf{k})] + \\ + \omega^- \cdot W(\omega/T_2, \omega^- \gamma/T_1) [\alpha_e''(\omega^- \gamma) \operatorname{Im} R_e^-(\omega,\mathbf{k}) + \alpha_m''(\omega^- \gamma) \operatorname{Im} R_m^-(\omega,\mathbf{k})] \end{cases} -$$

$$+ \frac{\hbar \gamma}{\pi^2} \iiint_{k<\omega/c} d\omega d^2 k \, \tilde{q}_0^{-1} (-\sin(2\tilde{q}_0 z)) \{R_e^\pm, R_m^\pm \to \tilde{R}_e^\pm, \tilde{R}_m^\pm\} -$$

$$+ \frac{\hbar \gamma}{\pi^2} \iiint_{k<\omega/c} d\omega d^2 k \, \tilde{q}_0^{-1} \cos(2\tilde{q}_0 z) \cdot$$

$$\cdot \begin{cases} \omega^+ \cdot W(\omega/T_2, \omega^+ \gamma/T_1) [\alpha_e''(\omega^+ \gamma) \operatorname{Re} \tilde{R}_e^+(\omega,\mathbf{k}) + \alpha_m''(\omega^+ \gamma) \operatorname{Re} \tilde{R}_m^+(\omega,\mathbf{k}] - \\ \omega^- \cdot W(\omega/T_2, \omega^- \gamma/T_1) [\alpha_e''(\omega^- \gamma) \operatorname{Re} \tilde{R}_e^-(\omega,\mathbf{k}) + \alpha_m''(\omega^- \gamma) \operatorname{Re} \tilde{R}_m^+(\omega,\mathbf{k}] \end{cases} \quad (14)$$

where $\hbar$, $k_B$ are the Planck's and Boltzmann's constants; the one and doubly primed functions denotes the corresponding real and imaginary parts,

$$\omega_1 = \omega \gamma (1 + \beta x), \quad \omega^\pm = \omega \pm k_x V \tag{15}$$

$$q_0 = (k^2 - \omega^2/c^2)^{1/2}, \quad \tilde{q}_0 = (\omega^2/c^2 - k^2)^{1/2}, \quad k^2 = k_x^2 + k_y^2$$
$$q = (k^2 - \varepsilon(\omega)\mu(\omega)\omega^2/c^2)^{1/2}, \quad \tilde{q} = (\varepsilon(\omega)\mu(\omega)\omega^2/c^2 - k^2)^{1/2} \tag{16}$$

$$\Delta_e(\omega) = \left(\frac{\varepsilon(\omega) q_0 - q}{\varepsilon(\omega) q_0 + q}\right), \quad \tilde{\Delta}_e(\omega) = \left(\frac{\varepsilon(\omega) \tilde{q}_0 - \tilde{q}}{\varepsilon(\omega) \tilde{q}_0 + \tilde{q}}\right) \tag{17}$$

$$\Delta_m(\omega) = \left(\frac{\mu(\omega) q_0 - q}{\mu(\omega) q_0 + q}\right), \quad \tilde{\Delta}_m(\omega) = \left(\frac{\mu(\omega) \tilde{q}_0 - \tilde{q}}{\mu(\omega) \tilde{q}_0 + \tilde{q}}\right) \tag{18}$$

$$\chi_e^{(\pm)}(\omega,\mathbf{k}) = 2(k^2 - k_x^2 \beta^2)(1 - \omega^2/k^2 c^2) + \frac{(\omega^\pm)^2}{c^2} \tag{19}$$

$$\chi_m^{(\pm)}(\omega,\mathbf{k}) = 2 k_y^2 \beta^2 (1 - \omega^2/k^2 c^2) + \frac{(\omega^\pm)^2}{c^2} \tag{20}$$

$$R_e^{(\pm)}(\omega,\mathbf{k}) = \chi_e^{(\pm)}(\omega,\mathbf{k}) \Delta_e(\omega) + \chi_m^{(\pm)}(\omega,\mathbf{k}) \Delta_m(\omega) \tag{21}$$

$$R_m^{(\pm)}(\omega,\mathbf{k}) = \chi_e^{(\pm)}(\omega,\mathbf{k}) \Delta_m(\omega) + \chi_m^{(\pm)}(\omega,\mathbf{k}) \Delta_e(\omega) \tag{22}$$

$$\tilde{R}_e^{(\pm)}(\omega,\mathbf{k}) = \chi_e^{(\pm)}(\omega,\mathbf{k}) \tilde{\Delta}_e(\omega) + \chi_m^{(\pm)}(\omega,\mathbf{k}) \tilde{\Delta}_m(\omega) \tag{23}$$

$$\tilde{R}_m^{(\pm)}(\omega,\mathbf{k}) = \chi_e^{(\pm)}(\omega,\mathbf{k}) \tilde{\Delta}_m(\omega) + \chi_m^{(\pm)}(\omega,\mathbf{k}) \tilde{\Delta}_e(\omega) \tag{24}$$

$$W(a/T_2, b/T_1) = \coth\left(\frac{\hbar a}{2 k_B T_2}\right) - \coth\left(\frac{\hbar b}{2 k_B T_1}\right) \tag{25}$$

In addition, it should be noted that all the integrals over the frequency $\omega$ are taken in the limits $(0,\infty)$, while over the projections $k_x, k_y$ of the planar wave vector **k** –in the first coordinate quadrant. The structure of the integrand functions within the brackets $\{R_e^\pm, R_m^\pm \to \tilde{R}_e^\pm, \tilde{R}_m^\pm\}$ is identical (with account of the replacement to be done) to the structure of the corresponding functions within the figure brackets of the integrals at $k > \omega/c$. In definitions frequently used by other authors, the coefficients $\Delta_e, \Delta_m, \tilde{\Delta}_e, \tilde{\Delta}_m$ coincide with the reflection factors for electromagnetic waves with different polarization [4, 9,10].

The first integral terms in (13), (14) describe interaction of a particle with vacuum background of temperature $T_2$ and do not depend on the distance to the surface $z$. For the first time, they have been obtained in [15] without magnetic polarization terms. The second and third terms (and both terms in (12)) depend on $z$ and describe interaction with the surface. In this case, the integrals computed over the domain of wave vectors $k > \omega/c$ are related with evanescent surface modes, while the integrals computed at $k < \omega/c$, are related with propagating surface modes.

Unlike the similar formulas which have been reported in [13,14], formulas (12) –(14) account for the contributions related with magnetic polarization of the particle, $\alpha_m$. On total, we see that Eqs. (12) -(14) manifest complete transposition symmetry over the electric (marked by the subscript "e") and magnetic (marked by the subscript "m") quantities. Quite recently, Eqs.(13), (14) have been obtained in [27, 28], and Eq.(12) (at $V = 0$) –in [16]. For neutral atoms formulas (12)-(13) must be taken at $T_1 = 0$. A heating of a neutral atom can be interpreted as some kind of Lamb –shift [7,8]. This question needs further theoretical elaboration.

## 5. Attraction force in the nonrelativistic case

In the nonrelativistic limit $\beta \ll 1, \gamma = 1$ formula (12) becomes simpler. Assuming the equilibrium case $T_1 = T_2 = T$, the result is [16]

$$F_z = -\frac{2\hbar}{\pi^2} \int_0^\infty d\omega \coth(\hbar\omega/2k_B T) \int_{k>\omega/c} d^2k \, \text{Im}\left( e^{-2q_0 z} [R_e(\omega,k)\alpha_e(\omega) + R_m(\omega,k)\alpha_m(\omega)] \right) - $$
$$-\frac{2\hbar}{\pi^2} \int_0^\infty d\omega \coth(\hbar\omega/2k_B T) \int_{k<\omega/c} d^2k \, \text{Im}\left( e^{2i\tilde{q}_0 z} [\tilde{R}_e(\omega,k)\alpha_e(\omega) + \tilde{R}_m(\omega,k)\alpha_m(\omega)] \right) \quad (26)$$

where $R_{e,m}(\omega,k)$ and $\tilde{R}_{e,m}(\omega,k)$ are given by (21)-(24) assuming $\beta = 0$ and making use the replacement $\Delta_{e,m}(\omega) \to \tilde{\Delta}_{e,m}(\omega)$. Using the identity $\coth(x) = 1 + 2/(\exp(x)-1)$, Eq.(26) can also be written as the sum of the zero point contribution $F_z(0,z)$ and thermal contribution $F_z(T,z)$:

$$F_z(z) = F_z(0,z) + F_z(T,z) =$$

$$-\frac{\hbar}{\pi}\int_0^\infty d\omega \left(\frac{\omega}{c}\right)^4 \int_1^\infty du\, u\, \exp(-2\omega z u/c) \cdot \left\{\begin{array}{l}\left[(2u^2-1)\tilde{\Delta}_e(u,\varepsilon) - \tilde{\Delta}_m(u,\varepsilon)\right]\alpha_e(i\omega) + \\ +\left[(2u^2-1)\tilde{\Delta}_m(u,\varepsilon) - \tilde{\Delta}_e(u,\varepsilon)\right]\alpha_m(i\omega)\end{array}\right\} -$$

$$-\frac{2\hbar}{\pi}\int_0^\infty d\omega \left(\frac{\omega}{c}\right)^4 \Pi(\omega,T)\int_0^1 du\, u\, \mathrm{Im}\left\{\exp(2i\omega z u/c)\left[\begin{array}{l}\left((1-2u^2)\tilde{\Delta}_e(u,\varepsilon) + \tilde{\Delta}_m(u,\varepsilon)\right)\alpha_e(\omega) + \\ +\left((1-2u^2)\tilde{\Delta}_m(u,\varepsilon) + \tilde{\Delta}_e(u,\varepsilon)\right)\alpha_m(\omega)\end{array}\right]\right\} - \quad (27)$$

$$-\frac{2\hbar}{\pi}\int_0^\infty d\omega \left(\frac{\omega}{c}\right)^4 \Pi(\omega,T)\int_0^\infty du\, u\, \mathrm{Im}\left\{\exp(-2\omega z u/c)\left[\begin{array}{l}\left((2u^2+1)\Delta_e(u,\varepsilon) + \Delta_m(u,\varepsilon)\right)\alpha_e(\omega) + \\ +\left((2u^2+1)\Delta_m(u,\varepsilon) + \Delta_e(u,\varepsilon)\right)\alpha_m(\omega)\end{array}\right]\right\}$$

where $\Pi(\omega,T) = \dfrac{1}{\exp(\hbar\omega/k_B T)-1}$, $\tilde{\Delta}_{e,m}(u,\varepsilon)$ in the first integral (27) is computed at imaginary frequency $i\omega$ with the substitution $u = \left(1 + \dfrac{k^2 c^2}{(i\omega)^2}\right)^{1/2}$; $\Delta_{e,m}(u,\varepsilon)$ and $\tilde{\Delta}_{e,m}(u,\varepsilon)$ in the second and third integrals (27) must be computed at real frequency $\omega$ making use the substitutions $u = \left(\dfrac{k^2 c^2}{\omega^2} - 1\right)^{1/2}$ and $u = \left(1 - \dfrac{k^2 c^2}{\omega^2}\right)^{1/2}$, respectively (see (17),(18)).

In the case $T_1 = T_2 = 0$ and $\alpha_m = 0$ formula (27) reduces to the "cold" Casimir (van –der – Waals) force between a particle and the surface, related to the zero –point fluctuations of electromagnetic field [29].

The thermal part of the Casimir force is given by the second and third terms of Eq. (27). The first one results from propagating modes of the surface, the second –from evanescent modes of the surface. In total, Eq.(27) represents the most general expression for the Casimir force between a resting particle and the surface at $R/z \ll 1$, which accounts for both electric and magnetic coupling, thermal and retardation effects.

It is worthwhile noticing that in metallic contacts the magnetic coupling effect is not small. Thus, in the case of ideal conductors at $T_1 = T_2 = 0$, it follows from (28) that the magnetic term equals $1/2$ of the electric one, and consequently, its contribution to the total Casimir force equals 30%. The corresponding force is given by $F_z(0,z) = -\dfrac{9\hbar c R^3}{4\pi z^5}$ in accordance with calculation from the first principles [30].

The principal non vanishing velocity expansion terms in Eq.(12) turn out to be proportional to $V^2$ and may be of two kinds. The first one behaves like $(V/\omega_0 z)^2$ in comparison with the zero temperature force (27) ($\omega_0$ is the characteristic absorption frequency), the second one behaves like $(V/c)^2$. Correspondingly, the former correction dominates in the nonretarded regime of interaction, at $z/\lambda < 1$, where $\lambda = c/\omega_0$ is the wave length of the absorption line. Typically, the dynamic corrections are small. However, at $\beta > z/\lambda$ they can be dominating.

## 5. Nonrelativistic dissipative (frictional) force

In what follows we assume $\mu = 1$ and the dielectric function of the surface is assumed to be $\varepsilon(\omega) = \varepsilon'(\omega) + i\varepsilon''(\omega)$. In the nonrelativistic limit $\beta \ll 1, \gamma = 1$ formula (13) is simplified, too. Expanding the integrand functions in (13) up to terms linear in velocity $V$ we get

$$F_x = F_x^{Vac} + F_x^S \tag{28}$$

$$F_x^{Vac} = -\frac{4\hbar V}{3\pi c^5} \int_0^\infty d\omega\, \omega^5 \cdot \left\{ \begin{array}{l} \dfrac{\hbar}{4k_B T_1}\left(\alpha_e'' + \alpha_m''\right)\sinh^{-2}\left(\hbar\omega/2k_B T_1\right) + \left[\Pi(\omega,T_2) - \Pi(\omega,T_1)\right] \cdot \\[6pt] \cdot\left[\dfrac{\alpha_e'' + \alpha_m''}{\omega} + \dfrac{1}{2}\dfrac{d\alpha_e''}{d\omega} + \dfrac{1}{2}\dfrac{d\alpha_m''}{d\omega}\right] \end{array} \right\} - \tag{29}$$

$$F_x^S = -\frac{\hbar V}{2\pi} \int_0^\infty \int_0^\infty d\omega\, du\, (\omega/c)^5 (u^2+1) \exp\left(-\frac{2\omega z}{c}u\right) \cdot$$

$$\cdot \left\{ \begin{array}{l} 2[\Pi(\omega,T_2) - \Pi(\omega,T_1)] \cdot \left[\dfrac{d\alpha_e''}{d\omega}\operatorname{Im} f_e + \dfrac{d\alpha_m''}{d\omega}\operatorname{Im} f_m + \dfrac{2}{\omega}(\alpha_e'' + \alpha_m'')(\Delta_e'' + \Delta_m'')\right] + \\[6pt] + \dfrac{\hbar}{2k_B T_1}\sinh^{-2}(\hbar\omega/2k_B T_1) \cdot \left[\alpha_e'' \cdot \operatorname{Im} f_e + \alpha_m'' \cdot \operatorname{Im} f_m\right] \end{array} \right\} -$$

$$-\frac{\hbar V}{2\pi} \int_0^\infty \int_0^1 d\omega\, du\, (\omega/c)^5 (1-u^2) \cdot$$

$$\cdot \left\{ \begin{array}{l} 2[\Pi(\omega,T_2) - \Pi(\omega,T_1)] \cdot \left[\operatorname{Re}\left(e^{\frac{2i\omega z}{c}u}\tilde{f}_e\right)\dfrac{d\alpha_e''}{d\omega} + \operatorname{Re}\left(e^{\frac{2i\omega z}{c}u}\tilde{f}_m\right)\dfrac{d\alpha_m''}{d\omega} + \right. \\[6pt] \left. + \dfrac{2}{\omega}\operatorname{Re}\left(e^{\frac{2i\omega z}{c}u}(\Delta_e + \Delta_m)\right)(\alpha_e'' + \alpha_m'') \right] + \\[6pt] + \dfrac{\hbar}{2k_B T_1}\sinh^{-2}(\hbar\omega/2k_B T_1) \cdot \left[\operatorname{Re}\left(e^{\frac{2i\omega z}{c}u}\tilde{f}_e\right)\alpha_e'' + \operatorname{Re}\left(e^{\frac{2i\omega z}{c}u}\tilde{f}_m\right)\alpha_m''\right] \end{array} \right\} \tag{30}$$

where the coefficients $f_{e,m}, \tilde{f}_{e,m}$ are given by

$$f_e = (2u^2 + 1)\Delta''_e(u,\varepsilon) + \Delta''_m(u,\varepsilon) \tag{31}$$

$$f_m = (2u^2 + 1)\Delta''_m(u,\varepsilon) + \Delta''_e(u,\varepsilon) \tag{32}$$

$$\tilde{f}_e = (1 - 2u^2)\tilde{\Delta}''_e(u,\varepsilon) + \tilde{\Delta}''_m(u,\varepsilon) \tag{33}$$

$$\tilde{f}_m = (1 - 2u^2)\tilde{\Delta}''_m(u,\varepsilon) + \tilde{\Delta}''_e(u,\varepsilon) \tag{34}$$

For brevity, the corresponding arguments of the functions in the integrands (27) are omitted. In Eq. (27) and (31)-(34) the imaginary components of the functions $\Delta_{e,m}(u,\varepsilon)$ and $\tilde{\Delta}_{e,m}(u,\varepsilon)$ must be computed from (17), (18) at real frequency with the substitution $u = \left(1 - \frac{k^2 c^2}{\omega^2}\right)^{1/2}$.

Eq.(29) represents the net vacuum component of the friction force, which does not depend on the distance to the surface. Eq.(30) represents the friction force due to the presence of the surface and, therefore, it depends on $z$. The first integral term (30) corresponds to the evanescent modes of the surface, the second –to the wave modes of the surface. At thermal equilibrium $T_1 = T_2 = T$ from (28)-(30) we get

$$F_x = -\frac{\hbar^2 V}{3\pi k_B T c^5}\int_0^\infty d\omega \frac{\omega^5}{\sinh^2(\hbar\omega/2k_B T)}(\alpha''_e + \alpha''_m) -$$
$$-\frac{\hbar^2 V}{4\pi k_B T c^5}\int_0^\infty\int_0^\infty d\omega du \, (u^2+1)\exp\left(-\frac{2\omega z}{c}u\right)\frac{\omega^5}{\sinh^2(\hbar\omega/2k_B T)}(\alpha''_e \operatorname{Im} f_e + \alpha''_m \operatorname{Im} f_m) - \tag{35}$$
$$-\frac{\hbar^2 V}{4\pi k_B T c^5}\int_0^\infty\int_0^1 d\omega du \, (1-u^2)\frac{\omega^5}{\sinh^2(\hbar\omega/2k_B T)}\left[\operatorname{Im}\left(e^{\frac{2i\omega z}{c}u}\tilde{f}_e\right)\alpha''_e + \operatorname{Im}\left(e^{\frac{2i\omega z}{c}u}\tilde{f}_m\right)\alpha''_m\right]$$

It is important that, contrary to the Casimir force (27) in the "cold" limit $T_1 = T_2 = 0$, the friction force proves to be zero. Eq.(35) is in agreement with the nonrelativistic results obtained by several authors [4,7,8,17,20,22,23], still generalizing them with account of magnetic effects (via the structure of the coefficients $f_{e,m}$, $\tilde{f}_{e,m}$ and $\alpha_m$), the vacuum and surface wave modes of electromagnetic field.

According to (29), (30), (35), and taking into account the relations $\alpha_e, \alpha_m \propto R^3$, it appears that the leading dependencies of the friction force components $F_x^{Vac}$ and $F_x^S$ vs. parameters $z, V, R, \omega_W$ are the following ($\omega_W = k_B T/\hbar$ implies the characteristic thermal (Wien) frequency) [27]:

i) vacuum component,

$$F_x^{Vac} \propto \frac{\hbar V \omega_W^2}{c^2}\left(\frac{\omega_W R}{c}\right)^3, \tag{36}$$

ii) surface component,

$$F_x^S \propto \begin{cases} -\dfrac{\hbar V}{z^2}\left(\dfrac{R}{z}\right)^3, & \omega_W z/c \ll 1 \\ -\dfrac{\hbar V}{z^2}\left(\dfrac{R}{z}\right)^3\left(\dfrac{\omega_W z}{c}\right)^3, & \omega_W z/c \gg 1 \end{cases} \qquad (37)$$

Comparing the nonretarded friction force (37) with the "cold" Casimir force $F_z(0,z) \propto -\dfrac{\hbar c R^3}{z^5}$ (the latter follows from (27)) we get $F_x^S/F_z \approx V/c \ll 1$ (assuming the nonrelativistic case ). In the ultrarelativistic case $\gamma \gg 1$ one should use Eqs.(12), (13) and the corresponding situation must be elaborated in more details.

As far as the force $F_x^{Vac}$ is concerned, the corresponding relativistic limit has been recently discussed in [15]. For a dielectric particle having a narrow absorption line with frequency $\omega_0$, one may write down $\alpha_e''(\omega) = 0.5\pi \omega_0 R^3 \delta(\omega-\omega_0)$, and the final result is

$$F_x^{Vac} = -\frac{\hbar \omega_0^2}{c\gamma^4}\left(\frac{\omega_0 R}{c}\right)^3 \cdot \int_{1/2}^{2\gamma^2} dx (1-x)[\Pi(\omega_0 x/\gamma, T_2) - \Pi(\omega_0, T_1)] \qquad (38)$$

Eq.(38) manifests a very intriguing possibility for the particle to be accelerated in the hot vacuum background, or even under the thermal equilibrium [15]. The involved dynamics equation $mcd\beta/dt = \gamma^{-3/2} F_x$ ($m$ is the particle mass) must be solved simultaneously with the equation for the particle temperature $T_1$ in its rest frame.

## 7. Nonrelativistic heating rate

The calculation of the particle heating rate results in the sum of the velocity independent heat flux, $dQ_0/dt$, and the dynamics corrections, of which the lowest–order one turns out to be proportional to $V^2$. It follows from (14),

$$\dot{Q}_0 = -\frac{4\hbar}{\pi c^3}\int_0^\infty d\omega \omega^4 \left[\alpha_e''(\omega)+\alpha_m''(\omega)\right]\cdot[\Pi(\omega,T_1)-\Pi(\omega,T_2)]-$$
$$-\frac{2\hbar}{\pi c^3}\int_0^\infty d\omega \omega^4 [\Pi(\omega,T_1)-\Pi(\omega,T_2)]\cdot\int_0^\infty du \exp\left(-\frac{2\omega z}{c}u\right)\cdot\left(\alpha_e'' \mathrm{Im}\, f_e + \alpha_m'' \mathrm{Im}\, f_m\right)- \qquad (39)$$
$$-\frac{2\hbar}{\pi c^3}\int_0^\infty d\omega \omega^4 [\Pi(\omega,T_1)-\Pi(\omega,T_2)]\cdot\int_0^1 du\left[\mathrm{Re}\left(e^{\frac{2i\omega z}{c}u}\tilde{f}_e\right)\alpha_e'' + \mathrm{Re}\left(e^{\frac{2i\omega z}{c}u}\tilde{f}_m\right)\alpha_m''\right]$$

Of course, $\dot{Q}_0 \neq 0$ only when $T_1 \neq T_2$. The first integral term of Eq. (39) is related with the net heat exchange between the particle and vacuum background and does not depend on $z$, the second is related with the evanescent wave modes of the surface and the third –with the surface wave modes, both depending on distance

The vacuum contribution to the heat flux can be simply obtained using the Kirchhgoff law of thermal radiation and the particle cross –section for the absorption of electromagnetic radiation [24]. However, the surface contributions to the heat flux can be worked out only on the basis of general theory of electromagnetic fluctuations. In the presented form, Eq.(39) has been firstly obtained in [29]. In comparison with the results reported by several authors [6,9,12,31], Eq. (39) incorporates both electric and magnetic effects of the particle polarization, and all contributions to the heat flux resulting from interactions with vacuum modes, evanescent and wave modes of the surface. On the contrary, the formulas given in [6,9,12,31] take into account only effects related with the evanescent surface modes, while the involved integrand functions do not yet contain important contributions resulting from the currents of magnetic polarization.

At $V \neq 0$, the process of radiative heat transfer manifests new features. Thus, the heat flux can be observed even at thermal equilibrium $T_1 = T_2 = T$. This is an obvious consequence of the energy conservation law: even if the dissipation integral in (7) equals zero, one still has a squared velocity term arising from work of friction force $-\mathbf{F}\cdot\mathbf{V}$. Another contribution results from Joule dissipation integral. The explicit formula can be obtained making use the velocity expansion in Eq. (14). The sign of the heat flux at thermal equilibrium may be different. In the case of small temperature difference $\Delta T$ between a particle and the surface, the velocity correction dominates, to a rough estimate, at $\beta^2 > \Delta T/T$.

In the ultrarelativistic limit $\gamma \gg 1$, at the same conditions as in sect. 6 ($\alpha_e'' \sim \delta(\omega-\omega_0)$), the vacuum background contributes to the heat flux, as follows [15]

$$\dot{Q} = \frac{\hbar\omega_0^5 R^3 \gamma^{-4}}{2c^3} \int_{1/2}^{2\gamma^2} dx \left(\Pi(\omega_0 x/\gamma, T_2) - \Pi(\omega_0, T_1)\right) \tag{40}$$

According to (38) and (40), the work of fluctuating electromagnetic field over the particle leads to its heating/cooling and slowing down (accelerating). The energy balance is given by

$$-dW/dt = F_x V + dQ/dt = \frac{\hbar\omega_0^5 R^3 \gamma^{-4}}{c^3} \int_{1/2}^{2\gamma^2} dxx \left(\Pi(\omega_0 x/\gamma, T) - \Pi(\omega_0, T)\right) > 0 \tag{41}$$

where $W$ is the electromagnetic field energy. This is in accordance with the general equation (7). The same behavior, of course, must be characteristic for the process of FEI regarding the surface modes.

## 8. Some numerical estimations
### i) Damping of an AFM tip

A development of the spectroscopy of dissipative (friction) forces in dynamic vacuum contacts of the AFM tips with surfaces and an extraction of meaningful physical information from the measurements are of great importance. A crucial factor in the interpretation of the experiments is that, in the noncontact dynamic regime with compensation for the contact potential difference, the conservative and dissipative interaction of a tip with the sample is expected to be associated with the forces of FEI (predominantly, of van –der –Waals type at nanometer separations). However, the theoretical calculations made by several authors have demonstrated that the vacuum friction is smaller than the observed friction by many orders of magnitude (see [7-9] and references there in). This enforces considering special mechanisms and conditions for the tip damping [9].

Here we propose one possible explanation for the damping effect in line with the obtained theoretical results. Let us consider the case of two coinciding resonance absorption peaks characterizing the particle and surface, both locating at $\omega_0 = \omega_W t_0$, $t_0 \ll 1$, $\omega_W$ is the Wien frequency. Assuming the nonretarded regime of the interaction, a dominating contribution to the friction force (35) will be related with the second term. The corresponding structure of the integrand function is simplified if use is made of the relations which hold close to $\omega_0$:

$$\Delta_e'' \approx \text{Im} \frac{\varepsilon_s - 1}{\varepsilon_s + 1} = \frac{2\varepsilon_s''}{(\varepsilon_s' + 1)^2 + \varepsilon_s''^2} \approx \frac{2}{\varepsilon_s''}, \text{ and } \alpha_e'' = R^3 \text{Im} \frac{\varepsilon_t - 1}{\varepsilon_t + 2} = R^3 \frac{3\varepsilon_t''}{(\varepsilon_t' + 2)^2 + \varepsilon_t''^2} \approx \frac{3R^3}{\varepsilon_t''} \quad (42)$$

where $\varepsilon_t$ and $\varepsilon_s$ are the dielectric functions of materials (the tip and the surface). Then Eq.(35) reduces to

$$F_x \approx -\frac{9}{4\pi} \frac{\hbar V R^3}{z^5} \int_{t_0 - \Delta t/2}^{t_0 + \Delta t/2} \frac{dt}{\sinh^2(t/2)} \frac{1}{\varepsilon_t''(\omega_W t) \varepsilon_s''(\omega_W t)} = -\frac{9}{\pi} \frac{\hbar V R^3}{z^5} \frac{\omega_W}{\omega_0^2} \frac{\Delta \omega_0}{\varepsilon_s''(\omega_0) \varepsilon_t''(\omega_0)} \quad (43)$$

with $\Delta t / t_0 = \Delta \omega_0 / \omega_0$. In the typical experimental situation [32,33], assuming the tip cross-section to be parabolic with the curvature radius $R$ and the height $H$, Eq.(43) can be considered as local relation in the differential volume $d^3 r$. Then, substituting $R^3 \to 3 d^3 r / 4\pi$ in (43) and integrating over the tip volume at $H/R \gg 1$, we obtain [27]

$$F_x = -\frac{9}{8\pi} \frac{V}{z_0^2} \frac{R}{z_0} \frac{k_B T}{\omega_0^2} \frac{\Delta \omega_0}{\varepsilon_s''(\omega_0) \varepsilon_t''(\omega_0)} \quad (44)$$

where $z_0$ implies the minimal tip-sample distance and $R$ is the tip curvature radius. For $T = 300 K$, $R = 35\, nm$, $z_0 = 10\, nm$ (experimental conditions [32,33]), and assuming $\omega_0 = 10^9\, s^{-1}$, $\Delta \omega_0 / \omega_0 = 0.1$, $\varepsilon_s'' = \varepsilon_t'' = 0.01$, we get from (44) the frictional stress $5.2 \cdot 10^{-11}\, N \cdot s \cdot m^{-1}$, that is close to the experimental value $(3.5 \div 13.5) \cdot 10^{-11}\, N \cdot s \cdot m^{-1}$ [33] and agrees with the observed distance dependence $F_x \propto z_0^{-3}$. However, in this case the surface atomic species of the tip and sample (in [33] both are made of gold) must have large absorption in the radio frequency range. It is worthwhile noticing that vibrational transitions in molecular species are located in the radio frequency domain, and the same order of magnitude has the inverse time for the damping motion of adsorbates measured in friction experiments using the quartz crystal microbalance technique [34]. Moreover, the order of the inverse relaxation time ($\sim 10^9\,s$ in [34]) is characteristic for all the experiments [32 -35] when using a phenomenological friction model [36].

**ii) Friction forces in metallic contacts**

Now we aim to demonstrate large influence of the magnetic polarization effect on friction and heat exchange in vacuum contacts of nonmagnetic metals. Let us consider vacuum friction between a metallic particle and the metallic surface in the case of thermal equilibrium, when Eq.(35) is correct. First, we assume the nonretarded regime of interaction, $\omega_W z/c \ll 1$. The dielectric function of the particle and surface materials, and the corresponding electric (magnetic) polarizabilities are assumed to be [24]

$$\varepsilon(\omega) = 1 + \frac{4\pi\sigma_0 i}{\omega} \tag{45}$$

$$\alpha_e(\omega) = R^3 \frac{\varepsilon(\omega) - 1}{\varepsilon(\omega) + 2} \tag{46}$$

$$\alpha_m(\omega) = -\frac{R^3}{2}\left[1 - \frac{3}{R^2 k^2} + \frac{3}{Rk}\cot(Rk)\right] = -R^3 \varphi_m(Rk),\ k = (1+i)\sqrt{2\pi\sigma_0\omega}/c \tag{47}$$

where $\sigma_0$ implies the static conductivity.

The calculated electric (including only electric polarization) and magnetic (including only magnetic polarization) contributions to the friction force (35) are shown in Fig.2 (for a Cu –Cu contact). We have assumed $R = 50\,nm,\ T = 300K,\ \sigma_0 = 5.2\cdot 10^{17}\,s^{-1},\ V = 1cm/s$. The accepted value of $V$ corresponds to the characteristic maximal velocity of the AFM tip in dynamic regime [32,33]. As seen from Fig.2, at $z > 10\,nm$ the contribution due to magnetic coupling dominates the contribution from electric coupling by 5 to 10 orders of magnitude. However, the corresponding friction coefficient $\eta = F_x/V = 10^{-20}\,N\cdot s/m$ (at $z = 10\,nm$, $T = 900K$) is yet too small to explain the observed experimental damping [32,33,35], where $\eta = 10^{-11} \div 10^{-13}\,N\cdot s/m$. Lines 5,6 on Fig.2 correspond to the vacuum contribution, being independent on the surface separation.

Fig.3 shows the calculated friction forces in retarded regime for large copper particles (with $R = 10$ and $50\,\mu m$) near the surface of copper at different temperatures. In this case the contribution from electric polarization turns out to be negligibly small. The friction forces decay like $F_x \sim T^a / z$ with $a = 2 \div 2.2$.

For a dielectric particle having a resonance absorption line in the microwave region, the friction force near a metallic surface can be somewhat larger than for a metallic particle of the same radius. By the same way, as in deriving Eq.(43), we obtain [27]

$$F_x = -2.3 \frac{\hbar V}{z^2} \left(\frac{R}{z}\right)^3 \left(\frac{\omega_W z}{c}\right)^4 \frac{\sigma_0}{\omega_W} \frac{\Delta\omega_0}{\omega_0} \frac{1}{\varepsilon''(\omega_0)}, \quad \omega_0 = 2.6\omega_W \tag{48}$$

In this case, as it follows from Eq.(48), the friction force decays like $F_x \sim T^3 / z$.

**iii) Vacuum heat exchange in metallic contacts**

Due to "truncating" character of the temperature factors entering into the frequency integrals (39), the main contribution comes from the low frequency domain $\omega < \omega_W$. Taking into account (45),(46), we have $\alpha_e'' = 3R^3 \omega / 4\pi\sigma_0 \ll R^3$, since for normal metals $\omega_W / \sigma_0 \sim 10^{-5} \div 10^{-4}$. At the same time, the function $\varphi_m(x)$ in Eq.(47) has smooth maximum 0.03 at $x = 4.8$. This implies that the contribution of electric dipole moment to the heating rate of the particle is always significantly smaller than the contribution of magnetic moment.

Fig. 4 shows the calculated relative rate of cooling of copper particles with $R = 3$ and $50 nm$ vs. distance to the copper surface at elevated temperature (with allowance for the temperature dependence of resistivity [37]). The numerical data are normalized to the rate of cooling corresponding to the Stefan–Boltzmann law

$$\dot{Q}_{BB} = -\frac{\pi^3}{15} \left(\frac{R\omega_W}{c}\right)^2 \hbar\omega_W^2 \tag{49}$$

According to these results, a heated metal nanoparticle is a source of intense thermal radiation. Its radiation power is by 2-6 orders of magnitude larger than the radiation power from a hypothetical blackbody of the same size.

With neglect of magnetic polarizability and with no account of retardation effects, under the same conditions otherwise, Eq.(39) yields [6-9,17]

$$\dot{Q}_{ev} = -\frac{\pi}{40} \left(\frac{R}{z}\right)^3 \left(\frac{\omega_W}{\sigma_0}\right)^2 \hbar\omega_W^2 \tag{50}$$

A comparison of formulae (49) and (50) reveals that $\dot{Q}_{ev} / \dot{Q}_{BB} < 0.01$ even at $R = z$. Therefore, it is the presence of fluctuation magnetic moment of the particle that accounts for dominating contribution to the rate of heat exchange with metallic surface. Also, formula (50) implies that resistive materials like an amorphous carbon can produce an intensive thermal radiation if $\omega_W = \sigma_0$ [6]. However, a comparison of the corresponding radiation power to that of a metallic particle of the same size shows that metal nanoparticles with radii above $\sim 10 nm$ are more effective radiators. Radiation power of such particles decays slower with distance ($\dot{Q} \sim 1/z$) than for resistive nanoparticles ($\dot{Q}_{ev} \sim 1/z^3$). The use of resistive materials is advantageous only for very small particles ($R < 10 nm$) and small separations from the surface.

An analysis of the contribution due to the vacuum background (the first integral (39)) shows that it does not exceed the black body radiation power (49) even for large particles with radii up to several microns. But in total, in a wide range of distances from the metal surface, the heat flux from the micron–sized metal particle significantly dominates the heat flux emitted by a blackbody of the same radius (see Fig.6).

It is most interesting to compare the recently obtained experimental data on radiative heat transfer with theoretical estimates. Thus, the authors of [38] have reported on measurements of near field heat transfer between the tip of a scanning thermal microscope, being made of electrochemically etched Pt–Ir wire, and surfaces of gold (Au) and gallium nitride (GaN). Below we briefly discuss the results relevant to an Au substrate, because the involved theoretical interpretation is straightforward.

Our first estimate can be obtained using the data shown in Fig.4 and scaling rule $\dot{Q} \propto R^3$. Thus, extrapolating the data corresponding to line 2 on Fig.4b to the experimental tip radius $R = 60 nm$ [38], with account of Eq.(49) we get $\dot{Q} \approx 3.4 \cdot 10^{-7} W$ at a distance of closest tip approach. As long as $z << R$, the calculated dependence on $z$ proves to be weak: $\dot{Q} \propto (R+z)^{-1}$. The measured heat flux is larger, $\dot{Q} \approx 8 \cdot 10^{-6} W$ [38], but the involved distance dependence in the near field region (up to separations of several tens $nm$) exhibits an extensive plateau, too.

Second, we have recalculated the heat flux using material parameters of bulk $Au$ both for the tip and sample [39]: $\omega_p = 1.37 \cdot 10^{16} \ rad/s$ (plasma frequency) and $\gamma = 5.32 \cdot 10^{12} \ rad/s$ (relaxation frequency). In this case we have got $\dot{Q} \approx 5.3 \cdot 10^{-9} W$ at the nearest separation distance.

The estimates being much closer to the experimentally observed ones could be obtained if use is made of a parabolic tip model (see section 8(i)). Thus, for a heated Cu tip ($T = 300 K, R = 50 nm, H/R = 100$) above the cold surface of Cu we have got $\dot{Q} \approx 10^{-5} W$ [27]. In the case of a gold tip above the surface of gold (at $T = 300, R = 60 nm, H/R = 100$) the corresponding heat flux is estimated to be equal to $5.3 \cdot 10^{-7} W$.

Higher values of the heat flux in the case of copper are due to larger conductivity. It is worthwhile noting that a nearly constant heat flux in the near field regime directly comes out from the contribution related with magnetic polarization of the tip. The corresponding effect is explicitly determined by Eq.(39). Contrary to this, in order to explain the experimental results, the authors of [38] have used an artificial heuristic model with no allowance for the magnetic polarization effects.

## 9. Concluding remarks

For the first time, we present the most general results relevant to the problem of FEI relevant to a small moving nonmagnetic particle (an atom) and the surface of polarizable medium (dielectric or metal). The temperatures of the particle and the surface (and vacuum background) can be arbitrary assuming the process of interaction to be stationary. We obtain relativistic expressions for the components of fluctuation force (conservative and dissipative) and heating rate of a particle near the heated surface. The obtained expressions take into account the velocity, temperature, retardation effects and material properties of materials in the most general form. It is shown that fluctuation magnetic polarization of metal nanoparticle results in important contributions to FEI. In the calculations of the dissipative (frictional) force and rate of heating the magnetic contributions appear to be larger by several orders of magnitude than the corresponding contributions arising from electric polarization. We also present the nonrelativistic (but retarded) expressions for the attraction/friction force and heating rate.

Finally, we consider several numerical examples of the obtained results applied to the problems of AFM damping, friction and heating. In particular, we put forward a possible mechanism of the AFM damping which provides quantitative agreement with experimental data. The obtained theoretical estimates of the heat flux in the near field regime turn out to be also in reasonable quantitative agreement with the experimental measurements, demonstrating weak distance dependence of the near field heat flux.

## FIGURES

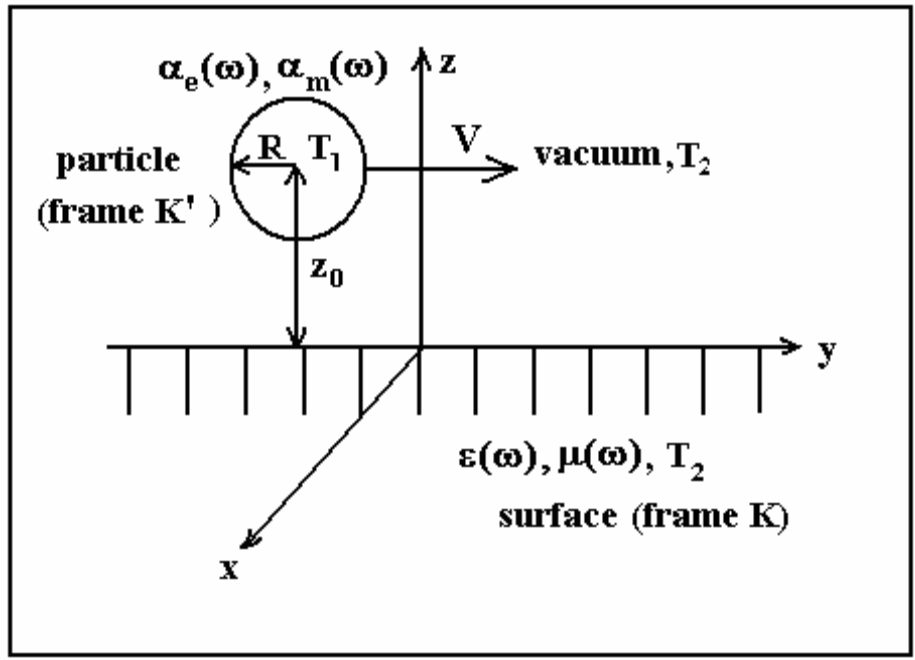

**Figure 1**

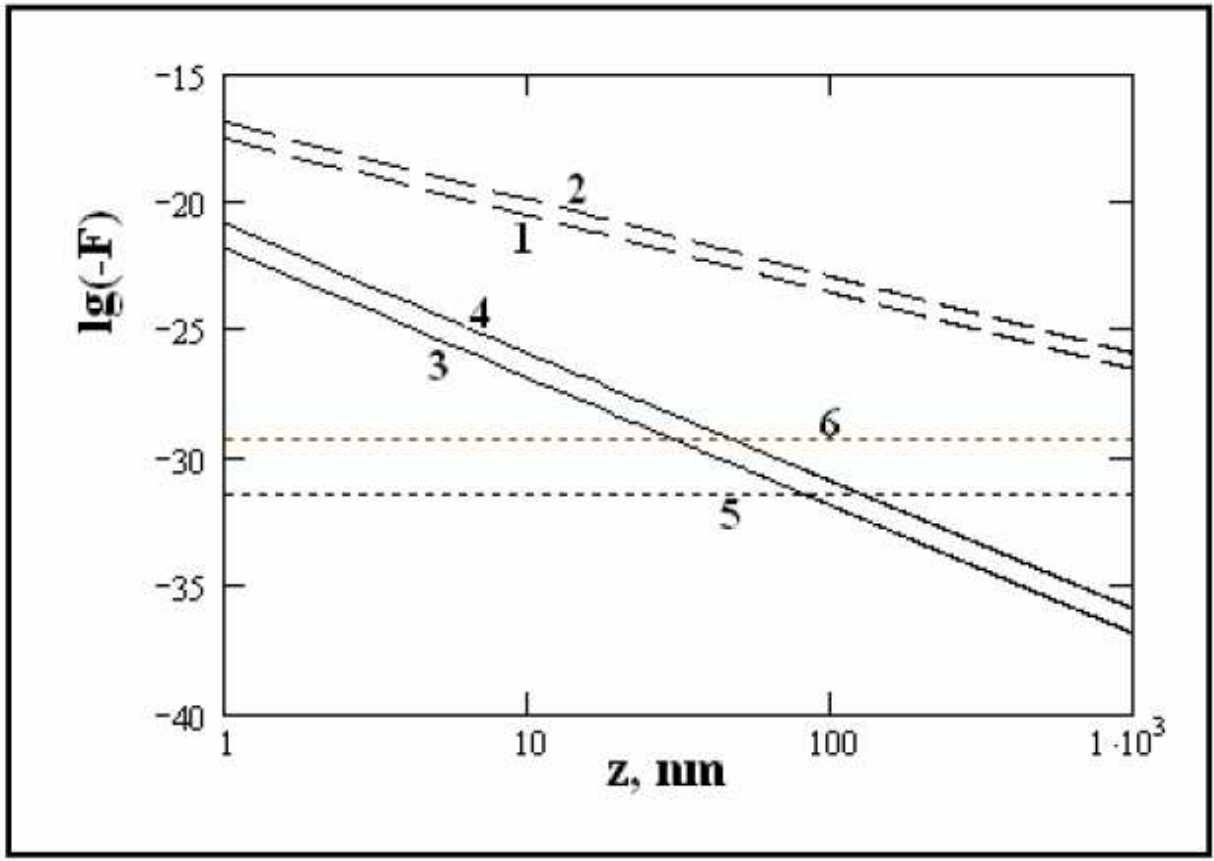

**Figure 2**

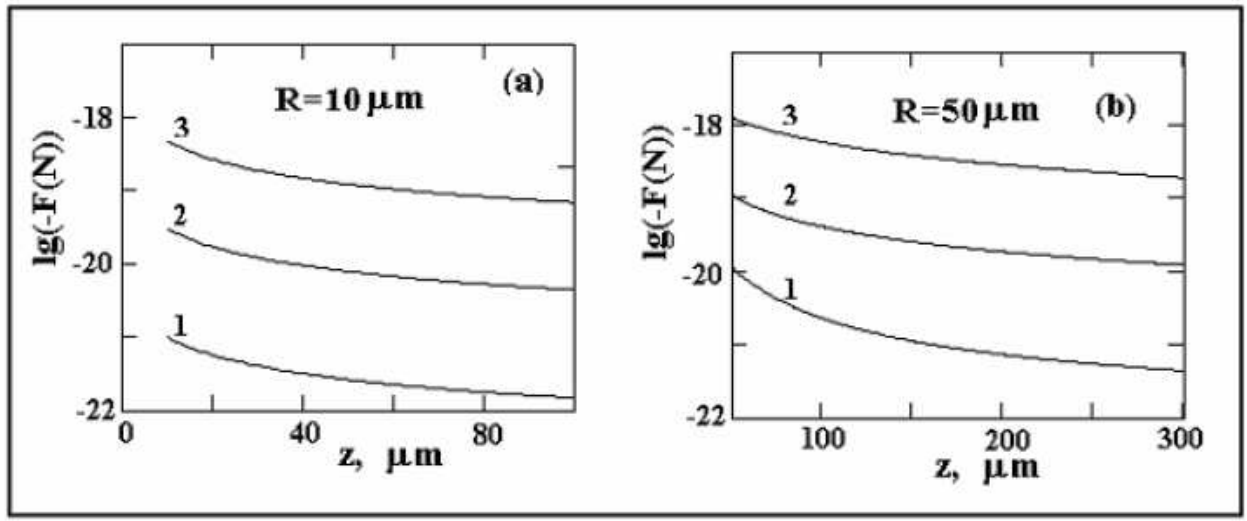

**Figure 3**

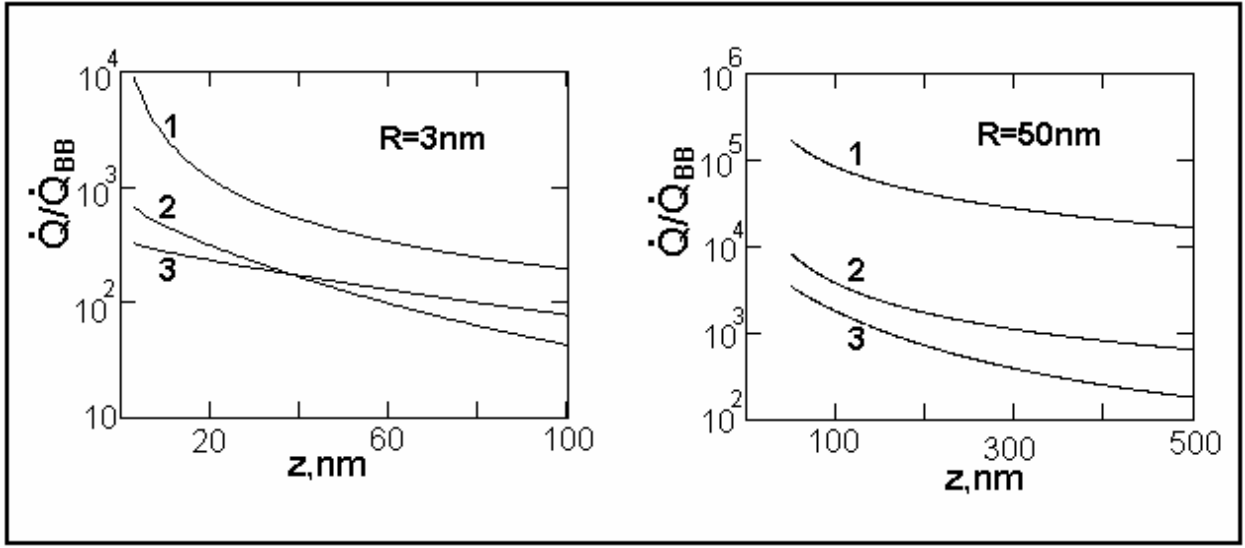

**Figure 4**

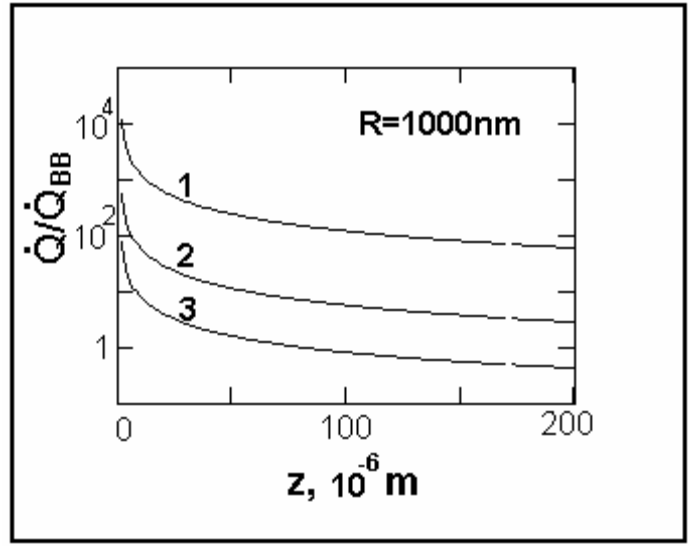

**Figure 5**

**Figure captions**

Fig.1
A moving particle over a half –space and coordinate systems used. The $x', y', z'$ coordinate axes are related with the particle reference frame $K'$ ( not pictured).

Fig.2
Plot of the sliding friction force (Eq.(35)) between a copper particle ( $R = 50\,nm, V = 1\,cm/s$ ) and the copper surface in dependence of distance and temperature. Lines 1,2 correspond to the contributions from only the magnetic polarization terms, lines 3, 4 – the contributions from electric polarization, lines 5,6 –the vacuum component of friction force (the first term in Eq.(35)). Lines 1,3,5 are computed at $T = 300K$ , lines 2,4,6 –at $T = 900K$ . The distance between the center of a spherical particle and the surface is $R + z$ .

Fig.3

The same as in Fig.3 for particles of radii 10 and 50microns at $V = 1m/s$. Lines 1,2,3 correspond to equilibrium temperatures $77K, 300K$ and $900K$, respectively. Here and in Figs.4,5 the distance $z$ is counted off the particle center.

Fig.4
Plots of the relative rate of radiative cooling (Eq.(39)) for copper nanoparticles with $R = 3nm$ (a) and $R = 50nm$ (b) and various temperatures $T_1 = 77(1)$, $300(2)$ and $900K(3)$ as a function of the distance to the copper surface; $Q_{BB}$ is the cooling rate for the blackbody of the same size. The temperature of vacuum background and the copper surface is $T_2 = 0K$

Fig.5
Same as in Fig.4 for a copper particle with $R = 1\mu m$.